\documentclass[aps,twocolumn,prb,eqsecnum,showpacs]{revtex4}
\usepackage{graphicx}
\begin{document}
\draft
\preprint{26 January 2006}
\title{Competing Ground States of Metal-Halide Ladders}
\author{Kei-ichi Funase and Shoji Yamamoto}
\address{Division of Physics, Hokkaido University,
         Sapporo 060-0810, Japan}
\date{26 January 2006}
\begin{abstract}
Based on a symmetry argument, we investigate the ground-state properties
of newly synthesized metal-halide ladder compounds
(C$_8$H$_6$N$_4$)[Pt(C$_2$H$_8$N$_2$)$X$]$_2$(ClO$_4$)$_4\cdot$2H$_2$O
($X=\mbox{Cl},\mbox{Br},\mbox{I}$).
Employing a fully dressed two-band Peierls-Hubbard model, we
systematically reveal possible charge- or spin-ordered states.
Numerical phase diagrams demonstrate a variety of competing Peierls and
Mott insulators with particular emphasis on the transition between two
types of mixed-valent state of Pt$^{\rm II}$ and Pt$^{\rm IV}$ driven by
varying interchain hopping integrals and Coulomb interactions.
\end{abstract}
\maketitle

\section{Introduction}

   The family of quasi-one-dimensional transition-metal ($M$) complexes
with bridging halogens ($X$) has been attracting much interest for several
decades. \cite{G6408,W6435}
They present an exciting stage performed by electron-electron correlation,
electron-lattice interaction, low dimensionality and $d$-$p$ orbital
hybridization, because their electronic state is widely tunable,
substituting the metals, halogens, ligand molecules and counterions.
The well-known Wolffram red salt
[Pt(ea)$_4$Cl]Cl$_2\cdot$2H$_2$O
($\mbox{ea}=\mbox{ethylamine}=\mbox{C}_2\mbox{H}_7\mbox{N}$) \cite{R2010}
possesses a Peierls-distorted mixed-valent ground state.
Its emission spectrum is so interesting as to consist of highly resonant
Raman lines and a remarkably Stokes-shifted luminescence band.
\cite{C287,T1446}
A nickel analog
[Ni(chxn)$_2$Br]Br$_2$
($\mbox{chxn}
 =\mbox{cyclohexanediamine}=\mbox{C}_6\mbox{H}_{14}\mbox{N}_2$)
\cite{T4261,T2341} has a Mott-insulating monovalent regular-chain
structure, but the bridging bromine ions can be photoexcited into
polarons. \cite{O465}
The competition between Peierls and Mott insulators \cite{N3865} is more
impressively observed in mixed-metal compounds
[Ni$_{1-x}$Pd$_x$(chxn)$_2$Br]Br$_2$ ($0\leq x\leq 1$).
\cite{M7699,W3948}
Metal binucleation stimulates further interest in the $M\!X$ family.
Two types of diplatinum-halide chain compounds,
 $A_4$[Pt$_2$(pop)$_4X$]$\cdot$$n$H$_2$O
($X=\mbox{Cl},\mbox{Br},\mbox{I}$;
 $A=\mbox{Li},\mbox{Cs},\cdots$;
 $\mbox{pop}=\mbox{diphosphonate}
 =\mbox{P}_2\mbox{O}_5\mbox{H}_2$) \cite{C4604,C409} and
 Pt$_2$($R$CS$_2$)$_4$I
($R=\mbox{alkyl\ chain}
 =\mbox{C}_n\mbox{H}_{2n+1}$) \cite{B444,I387}
exhibit mixed-valent ground states with halogen- and metal-sublattice
dimerization, respectively. \cite{B1155,K10068}
Photo- and pressure-induced phase transitions
\cite{M046401,Y140102,Y075113} are feasible in the former, while
successive phase transitions \cite{I387,K10068,Y1198} occur with
decreasing temperature in the latter.

   Several chemists \cite{Y} have recently made a brandnew attempt to
fabricate $M\!X$ ladder materials and obtained
(bpym)[Pt(en)$X$]$_2$(ClO$_4$)$_4\cdot$2H$_2$O
($X=\mbox{Cl},\mbox{Br},\mbox{I}$;
 $\mbox{en}=\mbox{ethylenediamine}=\mbox{C}_2\mbox{H}_8\mbox{N}_2$;
 $\mbox{bpym}=\mbox{bipyrimidine}=\mbox{C}_8\mbox{H}_6\mbox{N}_4$),
which comprise double-chain platinum complexes with intrachain bridging
halogen ions and interchain bridging bipyrimidine molecules.
A wider variety of mixed-valent states and their competition with Mott
insulators are expected of such a geometric crystalline structure.
SrCu$_2$O$_3$ \cite{H41} and NaV$_2$O$_5$ \cite{I1178} have also been
studied from a similar point of view.
The former behaves as a strongly correlated multiband $d$-$p$ ladder
especially under hole doping, \cite{U224406} while the latter is
describable within a single-band Hamiltonian \cite{S2602} but its $d_{xy}$
electrons are possibly coupled to phonons originating in the apical and/or
in-plane oxygen ions. \cite{R2667,S648}
The newly synthesized $M\!X$ ladder compounds are all the more fascinating
in that $d$-$p$ orbital hybridization and electron-lattice interaction are
both relevant.

   Thus motivated, we make our first attempt to reveal their ground-state
properties making group-theoretical analyses and Hartree-Fock
calculations.
A symmetry argument \cite{Y329,Y183,M2523} allows us to systematically
derive density-wave solutions and reduce the following numerical efforts
to the minimum necessary. \cite{Y125124}
A mean-field approximation applied to low-dimensional systems tends to
overstabilize broken-symmetry states against temperature
\cite{Y1198,K3825} but plays a leading role in revealing possible ground
states and interpreting their competition.
\cite{S2602,K1877,S3352,S1249,K1098,Y13}
Hartree-Fock calculations of $M\!M\!X$ chains \cite{Y125124} have indeed
characterized the pop- and $R$CS$_2$-ligand complexes as
$d_{z^2}$-single-band and $d_{z^2}$-$p_z$-hybridized two-band materials,
respectively, elucidating their distinct ground states of mixed valence.
Even the relaxation mechanism of photogenerated charge-transfer
excitations in mixed-valent $M\!X$ \cite{M5758,I1088} and $M\!M\!X$
\cite{O045122} chains have been calculated consistently with experimental
observations within the Hartree-Fock scheme.
The synthesized Pt$X$ ladders look double-chain analogs of the
conventional $M\!X$ chain compounds
[Pt(en)$_2X$](ClO$_4$)$_2$ and there are some similarities between their
intrachain absorption spectra. \cite{M}
The Hartree-Fock scheme combined with a symmetry argument promises to give
us a bird's-eye view of metal-halide ladders as well.

\section{Model Hamiltonian and Its Symmetry Properties}

\begin{figure}
\centering
\includegraphics[width=84mm]{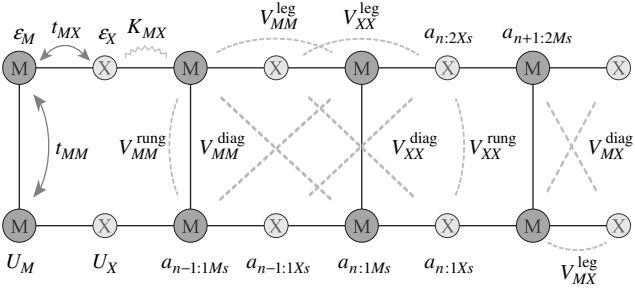}
\vspace*{-3mm}
\caption{The two-band extended Peierls-Hubbard modeling of the $M\!X$
         ladder.}
\label{F:H}
\end{figure}

   We consider a fully dressed three-quarter-filled two-band
Peierls-Hubbard Hamiltonian on the ladder lattice:
\begin{widetext}
\begin{eqnarray}
   &&
   {\cal H}
   =\sum_{n,l,s}
    \Bigl\{
     \bigl[\varepsilon_M-\beta_M(u_{n  :lX}-u_{n-1:lX})\bigr]
     n_{n:lMs}
    +\bigl[\varepsilon_X-\beta_X(u_{n+1:lM}-u_{n  :lM})\bigr]
     n_{n:lXs}
    \Bigr\}
   \nonumber\\
   &&\quad
   -\sum_{n,s}
    \Bigl\{
     \sum_{l}
     \bigl[t_{M\!X}+\alpha(u_{n:lM}-u_{n:lX})\bigr] 
     a_{n:lMs}^{\dagger}a_{n  :lXs}
    -\sum_{l}
     \bigl[t_{M\!X}-\alpha(u_{n:lM}-u_{n-1:lX})\bigr]
     a_{n:lMs}^{\dagger}a_{n-1:lXs}
   \nonumber\\
   &&\qquad\qquad
    +t_{M\!M}a_{n:1Ms}^{\dagger}a_{n:2Ms}
    +{\rm H.c.}
    \Bigr\}
   +\frac{K_{M\!X}}{2}\sum_{n,l}
    \bigl[(u_{n:lM}-u_{n:lX})^2+(u_{n:lM}-u_{n-1:lX})^2\bigr]
   \nonumber\\
   &&\quad
   +\sum_{n,l}\sum_{A=M,X}
    U_A n_{n:lA\uparrow}n_{n:lA\downarrow}
   +\sum_{n,l,s,s'}
    \Bigl\{
     V_{M\!X}^{\rm leg }n_{n:lMs}(n_{n:     l Xs'}+n_{n-1:     l Xs'})
    +V_{M\!X}^{\rm diag}n_{n:lMs}(n_{n:\bar{l}Xs'}+n_{n-1:\bar{l}Xs'})
    \Bigr\}
   \nonumber\\
   &&\quad
   +\sum_{n,s,s'}\sum_{A=M,X}
    \Bigl\{
     V_{AA}^{\rm rung}n_{n:1As}n_{n:2As'}
    +V_{AA}^{\rm diag}n_{n:1As}(n_{n-1:2As'}+n_{n+1:2As'})
    +\sum_{l}
     V_{AA}^{\rm leg}n_{n:lAs}n_{n-1:lAs'}
    \Bigr\},
   \label{E:H}
\end{eqnarray}
\end{widetext}
where $n_{n:lAs}=a_{n:lAs}^{\dagger}a_{n:lAs}$
($n=1,2,\cdots,N$; $l=1,2$; $A=M,X$; $s=\uparrow,\downarrow$)
with $a_{n:lAs}^{\dagger}$ creating an electron with spin $s$ for the
$M d_{z^2}$ ($A=M$) or $X p_z$ ($A=X$) orbital on the $l$th leg in the
$n$th unit, while $u_{n:lA}$ is the leg-direction displacement of the
metal ($A=M$) or halogen ($A=X$) on the $l$th leg in the $n$th unit from
its equilibrium position.
$\alpha$ and $\beta_A$ describe the Peierls- and Holstein-type
electron-lattice couplings, respectively, with $K_{M\!X}$ being the
metal-halogen spring constant.
Coulomb interactions are taken into consideration up to
next-nearest-neighbor metals and halogens, which is necessary and
sufficient considering that the intrachain and interchain metal spacings
are almost equal. \cite{Y}
Those between every pair of metals diagonally facing to each other,
$V_{M\!M}^{\rm diag}$, play a key role in controlling the interchain
charge arrangement.
A similar modeling has been applied to CuO $d$-$p$ ladders. \cite{N245109}
The notation is further explained in Fig. \ref{F:H}.

   When we consider normal states, the symmetry group of any lattice
electron system may be written as
$\mathbf{G}=\mathbf{P}\times\mathbf{S}\times\mathbf{T}$,
where $\mathbf{P}$, $\mathbf{S}$ and $\mathbf{T}$ are the groups of space,
spin rotation and time reversal, respectively.
For the present $d_{z^2}$-$p_z$ ladder $\mathbf{P}$ reads
$\mathbf{L}\land\mathbf{D}_2$, where
$\mathbf{D}_2=\{E,C_{2z},C_{2y},C_{2x}\}$ and ${\bf L}=\{E,l\}$ with $l$
being the one-dimensional translation by a unit cell.
Defining the Fourier transformation as
$a_{k:lMs}^\dagger
 =N^{-1/2}\sum_n{\rm e}^{{\rm i}kn}a_{n:lMs}^\dagger$ and
$a_{k:lXs}^\dagger
 =N^{-1/2}\sum_n{\rm e}^{{\rm i}k(n+1/2)}a_{n:lXs}^\dagger$ with the
lattice constant along the legs set equal to unity, we analyze all the
irreducible real representations of $\mathbf{G}$, which are
referred to as $\check{G}$, on the Hermitian-operator bases
$\{a_{k:lAs}^{\dagger}a_{k':l'A's'}\}$.
Actions of the space group elements on the electron operators are given in
Table \ref{T:GA}, whereas those of
$u(\mbox{\boldmath$e$},\theta)
 =\sigma^0\cos(\theta/2)
 -(\mbox{\boldmath$\sigma$}\cdot\mbox{\boldmath$e$})
  \sin(\theta/2)
 \in{\bf S}$
and $t\in{\bf T}$ are defined as
$u(\mbox{\boldmath$e$},\theta)\cdot a_{k:lAs}^\dagger
 =\sum_{s'}[u(\mbox{\boldmath$e$},\theta)]_{ss'}a_{k:lAs'}^\dagger$
and
$t\cdot a_{k:lAs}^\dagger
 =(-1)^{\delta_{s\uparrow}}a_{-k:lA-s}^\dagger$,
respectively, where $\sigma^0$ and
$\mbox{\boldmath$\sigma$}=(\sigma^x,\sigma^y,\sigma^z)$
are the $2\times 2$ unit matrix and a vector composed of the
Pauli matrices, respectively.
There is a one-to-one correspondence between $\check{G}$ and
broken-symmetry phases of density-wave type. \cite{O1514}
Any representation $\check{G}$ is obtained as a Kronecker product of
the irreducible real representations of ${\bf P}$, ${\bf S}$ and
${\bf T}$:
$\check{G}=\check{P}\otimes\check{S}\otimes\check{T}$.
$\check{P}$ is characterized by an ordering vector $q$ in the
Brillouin zone and an irreducible representation of its little group
${\bf P}(q)$, and is therefore labeled $q\check{P}(q)$.
The relevant representations of ${\bf S}$ are given by
$\check{S}^0(u(\mbox{\boldmath$e$},\theta))
 =1$ (nonmagnetic) and
$\check{S}^1(u(\mbox{\boldmath$e$},\theta))
 =O(u(\mbox{\boldmath$e$},\theta))$ (magnetic),
where $O(u(\mbox{\boldmath$e$},\theta))$ is the $3\times 3$
orthogonal matrix satisfying
$
   u(\mbox{\boldmath$e$},\theta)
   \mbox{\boldmath$\sigma$}^\lambda
   u^\dagger(\mbox{\boldmath$e$},\theta)
      =\sum_{\mu=x,y,z}
       [O(u(\mbox{\boldmath$e$},\theta))]_{\lambda\mu}
       \mbox{\boldmath$\sigma$}^\mu \ \
       (\lambda=x,\,y,\,z)
$,
whereas those of ${\bf T}$ by
$\check{T}^0(t)=1$ (symmetric) and $\check{T}^1(t)=-1$ (antisymmetric).
The representations
$\check{P}\otimes\check{S}^0\otimes\check{T}^0$,
$\check{P}\otimes\check{S}^1\otimes\check{T}^1$,
$\check{P}\otimes\check{S}^0\otimes\check{T}^1$ and 
$\check{P}\otimes\check{S}^1\otimes\check{T}^0$
correspond to charge-density-wave (CDW), spin-density-wave (SDW),
charge-current-wave (CCW) and spin-current-wave (SCW) states,
respectively.
\begin{table}[b]
\caption{The space group actions on the electron operators.}
\begin{ruledtabular}
\begin{tabular}{lcccc}
 & $l$ & $C_{2z}$ & $C_{2y}$ & $C_{2x}$ \\
\noalign{\vskip 1mm}
\hline
 $ a_{ k:1Ms}^{\dagger}$ & $e^{-{\rm i}k} a_{k:1Ms}^{\dagger}$ &
 $ a_{ k:2Ms}^{\dagger}$ & $a_{-k:1Ms}^{\dagger}$ &
 $ a_{-k:2Ms}^{\dagger}$ \\
 $ a_{ k:2Ms}^{\dagger}$ & $e^{-{\rm i}k} a_{k:2Ms}^{\dagger}$ &
 $ a_{ k:1Ms}^{\dagger}$ & $a_{-k:2Ms}^{\dagger}$ &
 $ a_{-k:1Ms}^{\dagger}$ \\
 $ a_{ k:1Xs}^{\dagger}$ & $e^{-{\rm i}k} a_{k:1Xs}^{\dagger}$ &
 $ a_{ k:2Xs}^{\dagger}$ & $-a_{-k:1Xs}^{\dagger}$ & 
 $-a_{-k:2Xs}^{\dagger}$ \\
 $ a_{ k:2Xs}^{\dagger}$ & $e^{-{\rm i}k} a_{k:2Xs}^{\dagger}$ &
 $ a_{ k:1Xs}^{\dagger}$ & $-a_{-k:2Xs}^{\dagger}$ &
 $-a_{-k:1Xs}^{\dagger}$ \\
 $a_{ k+\pi:1Ms}^{\dagger}$ & $-e^{-{\rm i}k} a_{k+\pi:1Ms}^{\dagger}$ &
 $a_{ k+\pi:2Ms}^{\dagger}$ & $a_{-k+\pi:1Ms}^{\dagger}$ &
 $a_{-k+\pi:2Ms}^{\dagger}$ \\
 $a_{ k+\pi:2Ms}^{\dagger}$ & $-e^{-{\rm i}k} a_{k+\pi:2Ms}^{\dagger}$ &
 $a_{ k+\pi:1Ms}^{\dagger}$ & $a_{-k+\pi:2Ms}^{\dagger}$ & 
 $a_{-k+\pi:1Ms}^{\dagger}$ \\
 $a_{ k+\pi:1Xs}^{\dagger}$ & $-e^{-{\rm i}k} a_{k+\pi:1Xs}^{\dagger}$ &
 $a_{ k+\pi:2Xs}^{\dagger}$ & $a_{-k+\pi:1Xs}^{\dagger}$ &
 $a_{-k+\pi:2Xs}^{\dagger}$ \\
 $a_{ k+\pi:2Xs}^{\dagger}$ & $-e^{-{\rm i}k} a_{k+\pi:2Xs}^{\dagger}$ &
 $a_{ k+\pi:1Xs}^{\dagger}$ & $a_{-k+\pi:2Xs}^{\dagger}$ &
 $a_{-k+\pi:1Xs}^{\dagger}$ \\
\end{tabular}
\end{ruledtabular}
\label{T:GA}
\end{table}
\begin{figure*}
\centering
\includegraphics[width=168mm]{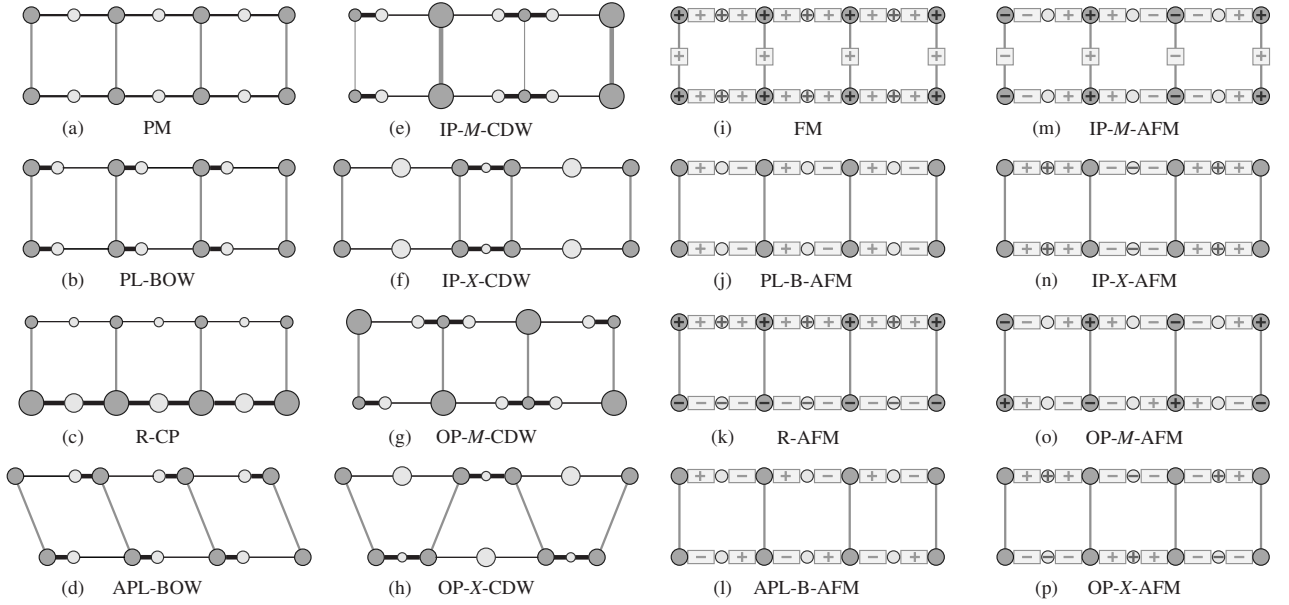}
\vspace*{-3mm}
\caption{Schematic representation of possible density-wave states,
         where the various circles and segments describe the variation of
         local charge densities
         $\sum_s\langle a_{n:lAs}^\dagger a_{n:lAs}\rangle_{\rm HF}$ and
         bond orders
         $\sum_s\langle a_{n:lAs}^\dagger a_{n':l'A's}\rangle_{\rm HF}$,
         respectively, whereas the signs $\pm$ in circles and strips
         denote the alternation of local spin densities
         $\sum_s\langle a_{n:lAs}^\dagger a_{n:lAs}\rangle_{\rm HF}
          \sigma_{ss}^z/2$ and
         spin bond orders
         $\sum_s\langle a_{n:lAs}^\dagger a_{n':l'A's}\rangle_{\rm HF}
          \sigma_{ss}^z/2$, respectively.
         Circles shifted from the regular position signify lattice
         distortion, which is peculiar to nonmagnetic phases.}
\label{F:DW}
\end{figure*}
\begin{table}[b]
\caption{Irreducible representations and characters for $\mathbf{D}_2$.}
\begin{ruledtabular}
\begin{tabular}{ccrrrr}
 $\mathbf{D}_2$ & Basis & $E$ & $C_{2z}$ & $C_{2y}$ & $C_{2x}$ \\
\noalign{\vskip 1mm}
\hline
 $A_1$ & $x^2,\,y^2,\,z^2$ & $ 1\,$ & $ 1\,$ & $ 1\,$ & $ 1\,$ \\
 $A_2$ & $xy$              & $ 1\,$ & $ 1\,$ & $-1\,$ & $-1\,$ \\
 $B_1$ & $xz$              & $ 1\,$ & $-1\,$ & $ 1\,$ & $-1\,$ \\
 $B_2$ & $yz$              & $ 1\,$ & $-1\,$ & $-1\,$ & $ 1\,$ \\
\end{tabular}
\end{ruledtabular}
\label{T:IC}
\end{table}

   We investigate static density waves of $q=0$ and $q=\pi$, which are
labeled $\mit\Gamma$ and $X$, respectively.
Then the Hamiltonian (\ref{E:H}) is rewritten within the Hartree-Fock
scheme as
\begin{eqnarray}
   &&
   {\cal H}_{\rm HF}
   =\sum_{lA,l'A'}
    \sum_{K={\mit\Gamma},X}
    \sum_{\lambda=0,x,y,z}
    \sum_{k,s,s'}
    x_{lAl'A'}^\lambda(K;k)
   \nonumber\\
   &&\qquad\quad\times
    a_{k+q(K):lAs}^\dagger a_{k:l'A's'}
    \sigma_{ss'}^\lambda,
   \label{E:HHF}
\end{eqnarray}
where $q({\mit\Gamma})=0$ and $q(X)=\pi$.
$x_{lAl'A'}^\lambda(K;k)$ can be expressed in terms of density matrices
$\rho_{l'A'lA}^\lambda(K;k)
 =\sum_{s,s'}\langle a_{k+q(K):lAs}^\dagger a_{k:l'A's'}\rangle_{\rm HF}
  \sigma_{ss'}^\lambda/2$,
where $\langle\cdots\rangle_{\rm HF}$ denotes the quantum average in a
Hartree-Fock eigenstate, and is determined self-consistently.
Since ${\bf P}({\mit\Gamma})={\bf P}(X)=\mathbf{D}_2$,
$\check{P}({\mit\Gamma})$ and $\check{P}(X)$ are both given by the
irreducible representations of $\mathbf{D}_2$, which are listed with their
characters in Table \ref{T:IC}.
Via group operations, the Hamiltonian (\ref{E:HHF}) is further
decomposed into symmetry-definite irreducible components as
\begin{equation}
   {\cal H}_{\rm HF}
   =\sum_{K={\mit\Gamma},X}
    \sum_{R=A_1,A_2,B_1,B_2}
    \sum_{\lambda=0,z}
    \sum_{\tau=0,1}
    h_{KR}^{\lambda\tau},
\end{equation}
where $h_{KR}^{\lambda\tau}$ is given by
\begin{eqnarray}
   &&
   h_{KR}^{\lambda\tau}
   =\frac{d^R}{2g}
    \sum_{t\in\mathbf{T}}\check{T}^\tau(t)
    \sum_{p\in\mathbf{D}_2}\chi^R(p)
    \sum_{lA,l'A'}\sum_{k,s,s'}tp
   \nonumber\\
   &&\qquad\quad
    \cdot
    x_{lAl'A'}^\lambda(K;k)
    a_{k+q(K):lAs}^\dagger a_{k:l'A's'}
    \sigma_{ss'}^\lambda.
\end{eqnarray}
$\chi^R(p)$ is the irreducible character of the $R$ representation for
the group element $p$, $g\,(=4)$ is the order of $\mathbf{D}_2$ and
$d^R\,(=1)$ is the dimension of $R$.
No representation of $d^R=1$ gives a helical SDW solution
($\lambda=x,y$).
The broken-symmetry Hamiltonians for
$KR\otimes\check{S}^0\otimes\check{T}^0$ and
$KR\otimes\check{S}^1\otimes\check{T}^1$
read
$h_{{\mit\Gamma}A_1}^{00}+h_{KR}^{00}$ and
$h_{{\mit\Gamma}A_1}^{00}+h_{KR}^{z1}$, respectively.
The lattice distortion $u_{n:lA}$ is determined so as to minimize the
Hartree-Fock energy $\langle{\cal H}\rangle_{\rm HF}$ and is also
described in terms of density matrices.
Considering that density matrices have the same symmetry properties
as their host Hamiltonian, we can qualitatively characterize all the
density-wave solutions, which are summarized in the following and
illustrated in Fig. \ref{F:DW}:

(a) ${\mit\Gamma}A_1\otimes\check{S}^0\otimes \check{T}^0$:
The paramagnetic state with the full symmetry $\mathbf{G}$,
abbreviated as PM.

(b) ${\mit\Gamma}A_2\otimes\check{S}^0\otimes\check{T}^0$:
Parallel bond order waves with the halogen sublattice distorted,
abbreviated as PL-BOW.

(c) ${\mit\Gamma}B_1\otimes\check{S}^0\otimes\check{T}^0$:
Polarized charge densities on rungs,
abbreviated as R-CP.

(d) ${\mit\Gamma}B_2\otimes\check{S}^0\otimes\check{T}^0$:
Antiparallel bond order waves with the halogen sublattice distorted,
abbreviated as APL-BOW.

(e) $XA_1\otimes\check{S}^0\otimes\check{T}^0$:
Charge density waves in phase on the metal sublattice with the halogen
sublattice distorted,
abbreviated as IP-$M$-CDW.

(f) $XA_2\otimes\check{S}^0\otimes\check{T}^0$:
Charge density waves in phase on the halogen sublattice with the metal
sublattice distorted,
abbreviated as IP-$X$-CDW.

(g) $XB_1\otimes\check{S}^0\otimes\check{T}^0$:
Charge density waves out of phase on the metal sublattice with the halogen
sublattice distorted,
abbreviated as OP-$M$-CDW.

(h) $XB_2\otimes\check{S}^0\otimes\check{T}^0$:
Charge density waves out of phase on the halogen sublattice with the metal
sublattice distorted,
abbreviated as OP-$X$-CDW.

(i) ${\mit\Gamma}A_1\otimes\check{S}^1\otimes\check{T}^1$:
Ferromagnetism on both metal and halogen sublattices,
abbreviated as FM.

(j) ${\mit\Gamma}A_2\otimes\check{S}^1\otimes\check{T}^1$:
Bond-centered antiferromagnetic spin densities on legs parallel to each
other,
abbreviated as PL-B-AFM.

(k) ${\mit\Gamma}B_1\otimes\check{S}^1\otimes\check{T}^1$:
Antiferromagnetic spin densities on rungs,
abbreviated as R-AFM.

(l) ${\mit\Gamma}B_2\otimes\check{S}^1\otimes\check{T}^1$:
Antiparallel spin bond order waves,
abbreviated as APL-B-AFM.

(m) $XA_1\otimes\check{S}^1\otimes\check{T}^1$:
Antiferromagnetic spin densities in phase on the metal sublattice,
abbreviated as IP-$M$-AFM.

(n) $XA_2\otimes\check{S}^1\otimes\check{T}^1$:
Antiferromagnetic spin densities in phase on the halogen sublattice,
abbreviated as IP-$X$-AFM.

(o) $XB_1\otimes\check{S}^1\otimes\check{T}^1$:
Antiferromagnetic spin densities out of phase on the metal sublattice,
abbreviated as OP-$M$-AFM.

(p) $XB_2\otimes\check{S}^1\otimes\check{T}^1$:
Antiferromagnetic spin densities out of phase on the halogen sublattice,
abbreviated as OP-$X$-AFM.

\section{Ground-State Phase Diagrams}

   The translationally dimerized phases (e)-(h) and (m)-(p) may be
expected to appear at zero temperature.
Pt$X$ chain compounds exhibit a ground state of the $M$-CDW type, while
Ni$X$ ones exhibit that of the $M$-AFM type.
When the chains are coupled in pairs, we wonder how their density waves
are stabilized in the ground state, in phase (IP) or out of phase (OP)
with each other.
We take another interest in whether or not a density wave may appear in
the $X$, rather than $M$, sublattice.
In order to visualize ground-state phase competition, we numerically
calculate $\langle{\cal H}\rangle_{\rm HF}$ for all the broken-symmetry
phases at a sufficiently low temperature ($k_{\rm B}T/t_{\rm M\!X}=0.05$)
in the thermodynamic limit ($N\rightarrow\infty$).
\begin{figure}
\centering
\includegraphics[width=84mm]{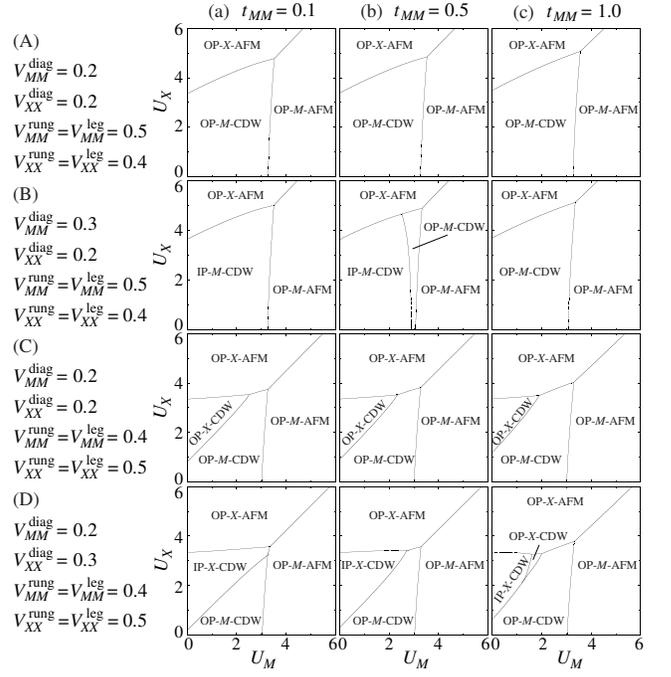}
\vspace*{-3mm}
\caption{Ground-state phase diagrams, where
         $\varepsilon_M-\varepsilon_X=1.0$,
         $\alpha=0.1$, $\beta_M=\beta_X=1.0$,
         $V_{M\!X}^{\rm leg}=1.0$ and $V_{M\!X}^{\rm diag}=0.4$ in common.}
\label{F:PhD}
\end{figure}

\begin{figure}
\centering
\includegraphics[width=84mm]{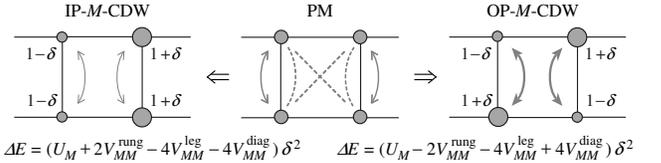}
\vspace*{-3mm}
\caption{A simple consideration of the competition between IP-$M$-CDW and
         OP-$M$-CDW.
         The Coulomb energy gains due to charge disproportionation,
         ${\mit\Delta}E$, are evaluated, provided the $X$ $p_z$ orbitals
         are fully filled and thus negligible.}
\label{F:IPvsOP}
\end{figure}

   Considering that the on-site Coulomb repulsion may vary with the
constituent metals and halogens as $U_{\rm Pt}<U_{\rm Pd}<U_{\rm Ni}$ and
$U_{\rm I}<U_{\rm Br}<U_{\rm Cl}$, we draw several ground-state phase
diagrams on the $U_M$-$U_X$ plane in Fig. \ref{F:PhD}, where the
electronic and phononic energies are scaled by $t_{M\!X}$ and $K_{M\!X}$,
respectively.
A variety of Peierls and Mott insulators are stabilized at moderate and
strong Coulomb interactions, respectively.
Charge or spin densities oscillate in the $M$ sublattice for
$U_X\alt U_M$, whereas in the $X$ sublattice for $U_M\alt U_X$.
AFM phases of the IP type are hardly stabilized, while CDW phases of the
IP and OP types closely compete with each other with varying
$V_{AA}^{\rm rung}$, $V_{AA}^{\rm diag}$ and $t_{M\!M}$.
The present AFM phase boundaries should be refined beyond the Hartree-Fock
scheme when Ni$X$ ladders are synthesized, where quantum correlations are
supposed to be significant. \cite{O1571}
Any AFM phase might be stabilized with weak but nonnegligible interactions
between neighboring ladders.

   Figures \ref{F:PhD}(A) and \ref{F:PhD}(B) demonstrate the competition
between the most likely phases IP-$M$-CDW and OP-$M$-CDW on condition that
the inter-metal Coulomb interactions are slightly larger than the
inter-halogen ones, which is understandable within the naivest
consideration of $d$-electron correlations.
Figure \ref{F:IPvsOP} shows that the two states may be balanced at
$V_{M\!M}^{\rm rung}\simeq 2V_{M\!M}^{\rm diag}$.
Indeed, there appears
OP-$M$-CDW at $2V_{M\!M}^{\rm diag}=0.4<0.5=V_{M\!M}^{\rm rung}$, while
IP-$M$-CDW at $2V_{M\!M}^{\rm diag}=0.6>0.5=V_{M\!M}^{\rm rung}$.
It is also interesting to observe the phase competition from the viewpoint
of electron itinerancy.
While IP-$M$-CDW and OP-$M$-CDW are degenerate to each other in decoupled
chains, the interchain electron hopping is much more advantageous to
OP-$M$-CDW.
Once $t_{M\!M}$ is on, IP-$M$-CDW and OP-$M$-CDW are both stabilized by an
energy $\propto t_{M\!M}^2$ in general, but the latter gains much more.
If we think of the limit of $\delta=1$, IP-$M$-CDW gains no energy with
increasing $t_{M\!M}$.
That is why IP-$M$-CDW is replaced by OP-$M$-CDW in Fig. \ref{F:PhD}(B-c).

   Figures \ref{F:PhD}(C) and \ref{F:PhD}(D) reveal a
possibility of the novel phases IP-$X$-CDW and OP-$X$-CDW appearing on
condition that the inter-metal Coulomb interactions are slightly smaller
than the inter-halogen ones.
IP-$X$-CDW and OP-$X$-CDW also compete with each other, the former of
which is stabilized by $V_{X\!X}^{\rm diag}$, while the latter of which
by $V_{X\!X}^{\rm rung}$.
The hybridization-driven phase competition was predicted for single chains
\cite{Y125124,R3498,Y422,K435} as well and was indeed observed in an
$M\!M\!X$ compound. \cite{K10068}
Since $U_X$ acts on the oxidation of $X^-$ ions, it is interesting to tune
PtCl ladders chemically, with varying ligands, for example, and
physically, by applying pressure, for instance.

\section{Summary}

   Charge- or spin-ordered states of the $M\!X$ ladder are thus various
and highly competing.
Raman spectra for (bpym)[Pt(en)$X$]$_2$(ClO$_4$)$_4\cdot$2H$_2$O suggest a
mixed-valent ground state of Pt$^{\rm II}$ and Pt$^{\rm IV}$, \cite{M} we
are for the moment encouraged to identify it as either IP-$M$-CDW or
OP-$M$-CDW.
In any case, it is highly interesting to apply pressure to the materials.
Sample size reduction along the rungs enhances $t_{M\!M}$, whereas that
along the legs effectively reduces $t_{M\!M}$.
Our findings claim that there is a good possibility of pressure-induced
phase transitions between IP-$M$-CDW and OP-$M$-CDW.

   Further phase transitions may be expected at finite temperatures and/or
under doping.
Photoexcitations and their nonlinear relaxation processes must be the
following interest.
We hope the present study will stimulate and guide further explorations
into the brandnew $M\!X$ materials.

\acknowledgments

   The authors thank K. Iwano, H. Okamoto and M. Yamashita for fruitful
discussions.
This work was supported by the Ministry of Education, Culture, Sports,
Science and Technology of Japan.

\end{document}